\documentclass[pre,aps,twocolumn,superscriptaddress,showpacs]{revtex4-1}

\usepackage{latexsym}
\usepackage{amssymb}
\usepackage{graphicx}  
\usepackage{epsfig}    
\usepackage{dcolumn}   
\usepackage{bm}        
\usepackage{color}     
\usepackage{ulem}      
\usepackage{url}       
\newcommand{\Epsilon}{\mathcal{E}}

\newcommand{\bx}{\mathbf{x}}

\begin{document}
\title[]{Coulomb explosion of uniformly charged spheroids}

\author{M.~Grech}
\email[]{mickael.grech@gmail.com}
\address{Max-Planck-Institute for the Physics of Complex Systems, D-01187 Dresden, Germany}

\author{R.~Nuter}
\address{CEA, DAM, DIF, F-91297 Arpajon, France}

\author{A.~Mikaberidze}
\address{Max-Planck-Institute for the Physics of Complex Systems, D-01187 Dresden, Germany}

\author{P.~Di~Cintio}
\address{Max-Planck-Institute for the Physics of Complex Systems, D-01187 Dresden, Germany}

\author{L.~Gremillet}
\address{CEA, DAM, DIF, F-91297 Arpajon, France}

\author{E.~Lefebvre}
\address{CEA, DAM, DIF, F-91297 Arpajon, France}

\author{U.~Saalmann}
\address{Max-Planck-Institute for the Physics of Complex Systems, D-01187 Dresden, Germany}

\author{J.~M.~Rost}
\address{Max-Planck-Institute for the Physics of Complex Systems, D-01187 Dresden, Germany}

\author{S.~Skupin}
\address{Max-Planck-Institute for the Physics of Complex Systems, D-01187 Dresden, Germany}
\address{Institute of Condensed Matter Theory and Solid State Optics, Friedrich-Schiller-University Jena, D-07743 Jena, Germany} 

\begin{abstract}
A simple, semi-analytical model is proposed for non-relativistic Coulomb explosion of a uniformly charged spheroid. This model allows us to derive the time-dependent particle energy distributions. Simple expressions are also given for the characteristic explosion time and maximum particle energies in the limits of extreme prolate and oblate spheroids as well as for the sphere. Results of particle simulations are found to be in remarkably good agreement with the model. 
\end{abstract}

      
\maketitle

\section{Introduction}\label{sec1}

Coulomb explosion (CE) is an ubiquitous phenomenon in laser-matter interaction, from laser ablation and micromachining to particle acceleration~\cite{nishihara_NIMA_2001, sakabe_PRA_2004, last_JCP_2004, islam_PRA_2006, kaplan_PRL_2003, kovalev_JETP_2005, esirkepov_PRL_2002, fourkal_PRE_2005, bulanov_PRE_2008, grech_LinPA, tikhonchuk_NIMA_2010, hashida_OE_2009}. CE is the dominant process of ion acceleration from a cluster irradiated by an intense laser pulse in the regime of so-called cluster vertical ionization (CVI)~\cite{nishihara_NIMA_2001, sakabe_PRA_2004, last_JCP_2004, islam_PRA_2006, kaplan_PRL_2003, kovalev_JETP_2005}. In this regime, the laser pulse is intense enough to remove all electrons from the cluster before ion motion sets in. This kind of ion charge state can also be generated by intense and short pulses of high energy photons which have become possible at x-ray free electron laser~\cite{saalmann_JPB_2006}. In both cases, the ion dynamics is governed by CE.

Spherical CE has been thoroughly investigated in the last years due to its importance for cluster physics. In the case of a uniformly charged sphere, CE is self-similar and can be described analytically~\cite{nishihara_NIMA_2001, sakabe_PRA_2004, last_JCP_2004, islam_PRA_2006}. The dynamics of CE of a non-uniformly charged sphere is more complex as it involves multiple flows so that a kinetic description is required~\cite{kaplan_PRL_2003, kovalev_JETP_2005}. 

In contrast, ellipsoidal and spheroidal (ellipsoidal with a rotational symmetry) CE has been studied in the context of accelerator physics, where three-dimensional (3D) envelope equations are widely used~\cite{davidson_2001,batygin_POP_2001}, or to model space charge effects in laser-created dense electron beams~\cite{fubiani_PRSTAB_2006}.

Spheroidal clusters have also attracted a lot of attention as they exhibit characteristic electron momentum distributions~\cite{rigo_PRB_1998}, and due to their optical properties which are of great interest in, e.g., nano-optics~\cite{dellafiore_PRB_2000, kelly_JPCB_2003}. Moreover, spheroidal clusters appear as a natural candidate for anisotropic ion emission from clusters under intense ultrashort laser irradiation, which has recently triggered significant interest in anisotropic cluster expansion~\cite{skopalova_PRL_2010}. Furthermore, it has also been shown that in helium embedded rare gas clusters, a spheroidal nanoplasma is generated by illumination with an intense laser pulse giving to unusual resonant heating~\cite{mikaberidze_PRL_2009}.

In addition, understanding spheroidal CE is crucial in the context of ion acceleration from a solid target irradiated by an intense, relativistic laser pulse. Recent studies have shown CE of thin multispecies target as a promising path toward high-quality ion beams~\cite{esirkepov_PRL_2002, fourkal_PRE_2005, bulanov_PRE_2008, grech_LinPA}. 
Apart from the possibility to use CE as the principal acceleration mechanism, CE of the accelerated ion bunch itself has been shown to play a dominant role in angular as well as energy dispersion of ion beams generated from laser-solid interaction~\cite{grech_LinPA, tikhonchuk_NIMA_2010}.

In this paper, we investigate CE of an initially uniformly charged spheroid. In order to derive simple estimates for the particle maximum energies and characteristic explosion time,  we restrict ourselves to non-relativistic particle velocities. We then demonstrate that, during CE, both the spheroidal shape and uniformity of the charge distribution are conserved, but with time-dependent aspect-ratio and charge density. Therefore CE of a uniformly charged spheroid can be described using a simple, semi-analytical model for the evolution of the spheroid radii. This model allows us to derive the temporal evolution of the particle energy distribution and maximum energies (along the spheroid principal axes) as a function of the spheroid initial aspect-ratio, charge density and total charge. Our theoretical predictions are then compared to molecular dynamics (MD) and 3D particle-in-cell (PIC) simulations. These simulation tools are the most widely used methods to model laser-cluster and laser-plasma interaction. However, they are known to be computationally costly, so that the results obtained in this paper are interesting for various applications, from non-spherical cluster CE to non-neutralized charged particle beam propagation through a vacuum.

The paper is structured as follows. Section~\ref{sec2} presents our semi-analytical model. Predictions from the model are then compared to both MD simulations (Sec.~\ref{sec3}) and PIC simulations (Sec.~\ref{sec4}). Finally, Sec.~\ref{sec5} summarizes our findings.


\begin{figure}
\begin{center}
\includegraphics[width=8cm]{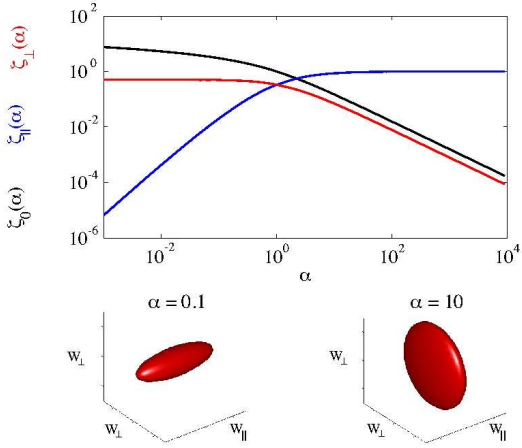}
\caption{Dependence of the shape-functions $\zeta_0(\alpha)$ (black curve), $\zeta_{\Vert}(\alpha)$ (blue curve) and $\zeta_{\perp}(\alpha)$ (red curve) on the aspect-ratio $\alpha$. The lower row shows a prolate (cigar-shaped) spheroid with $\alpha=0.1$ and an oblate (disk-shaped) spheroid with $\alpha=10$.}
\label{fig_zeta}
\end{center}
\end{figure}

\section{Semi-analytical model for Coulomb explosion of a uniformly charged spheroid}\label{sec2}

\subsection{General considerations on uniformly charged spheroids}\label{sec2.1}

In this first Section, we lay the basis for our simple model of CE of a uniformly charged spheroid. To do so, let us first recall the electrostatic potential at a position $\bx = (x,y,z)$ inside a uniformly charged ellipsoid centered in $\bx = 0$ and with radii $w_x$, $w_y$ and $w_z$ along the directions $x$, $y$ and $z$, respectively~\cite{landau_v2}:
\begin{widetext}
\begin{eqnarray}\label{eq_ellipsoidpotential}
\hspace{-2cm}\phi(\mathbf{x}) = \pi k\,\rho_c\,w_x\,w_y\,w_z\,\int_{0}^{\infty} \left(1-\frac{x^2}{w_x^2+s}-\frac{y^2}{w_y^2+s}-\frac{z^2}{w_z^2+s}  \right)\,\frac{ds}{\sqrt{\psi(s)}}\,,
\end{eqnarray}
\end{widetext}
where $k=(4\pi\epsilon_0)^{-1}$ and $\rho_c=Z\,e\,n$ is the charge density (typically $Z$ is the mean ion charge state and $n$ is the ion density), and $\psi(s) = (w_x^2+s)\,(w_y^2+s)\,(w_z^2+s)$. 

Objects with a rotational symmetry are of particular importance for many applications such as cluster explosion or particle acceleration. Hence, we introduce the radial coordinate $r = \sqrt{y^2+z^2}$ and restrict our study to the case of a spheroid: $w_x = w_{\Vert}$ and $w_y = w_z = w_{\perp}$ so that $\psi(s) = (w_{\Vert}^2+s)\,(w_{\perp}^2+s)^2$. Finally, one obtains for the electrostatic potential inside the spheroid:
\begin{eqnarray}\label{eq_spheroidpotential}
\phi(x,r) = 2\pi k\,\rho_c\,\left(w_{\perp}^2\,\zeta_0(\alpha)-\zeta_{\Vert}(\alpha)\,x^2 - \zeta_{\perp}(\alpha)\,r^2\right) ,
\end{eqnarray}
where we have introduced the spheroid aspect-ratio $\alpha=w_{\perp}/w_{\Vert}$ and:
\begin{eqnarray}\label{eq_shp_fct1}
\zeta_0(\alpha)       &=& \frac{1}{2}\,\int_0^{\infty} \frac{ds}{(\alpha^2+s)\,\sqrt{1+s}}\,,\\
\zeta_{\Vert}(\alpha) &=& \frac{\alpha^2}{2}\,\int_0^{\infty} \frac{ds}{(\alpha^2+s)\,(1+s)^{3/2}}\,,\\\label{eq_shp_fct3}
\zeta_{\perp}(\alpha) &=& \frac{\alpha^2}{2}\,\int_0^{\infty} \frac{ds}{(\alpha^2+s)^2\,\sqrt{1+s}}\,.
\end{eqnarray}
In what follows, we refer to these functions as shape-functions as they depend only on the spheroid aspect-ratio. Equations~(\ref{eq_shp_fct1})-(\ref{eq_shp_fct3}) are here written in their compact, integral forms. These integrals can, however, be expressed as functions of inverse trigonometric and hyperbolic functions, see e.g. Ref.~\cite{kiwamoto_POP_2004}. The dependencies of these shape-functions on $\alpha$ are given in Fig.~\ref{fig_zeta}. We also give their limits for the extreme prolate (cigar-shape, $\alpha \ll 1$) and  oblate (disc-shape, $\alpha \gg 1$) spheroids, as well as for the sphere ($\alpha=1$):
\begin{center}
\begin{tabular}{|c|c|c|c|}\hline
                         &  $\alpha \ll 1$        &  $\alpha = 1$   &  $\alpha \gg 1$       \\ \hline
$\zeta_0(\alpha)$        &  $\ln (2/\alpha)$              &  $1$            &  $\pi/(2\alpha)$      \\ \hline
$\zeta_{\Vert}(\alpha)$  &  $\alpha^2\,\big[\ln (2/\alpha)-1\big]$ &  $1/3$          &  $1$                  \\ \hline
$\zeta_{\perp}(\alpha)$  &  $1/2$                 &  $1/3$          &  $\pi/(4\alpha)$      \\ \hline
\end{tabular}
\end{center}

Equation~(\ref{eq_spheroidpotential}) illustrates the well-known result that the electrostatic potential inside a uniformly charged spheroid is a quadratic function of the space coordinates. As for the electric field inside the uniformly charged spheroid, it can be easily expressed in cylindrical coordinates:
\begin{eqnarray}
\mathbf{E}(x,r) = E_{\Vert}(x)\,\hat{\bx} + E_{\perp}(r)\,\hat{\mathbf{r}}\,,
\end{eqnarray}
where $\hat{\bx}$ and $\hat{\mathbf{r}}$ are the longitudinal and radial unit vectors, respectively, and:
\begin{eqnarray}
\label{eq_field_1} E_{\Vert}(x)=-\partial_x\,\phi(x,r)=4\pi k\,\rho_c\,\zeta_{\Vert}(\alpha)\,x\,,\\
\label{eq_field_2} E_{\perp}(r)=-\partial_r\,\phi(x,r)=4\pi k\,\rho_c\,\zeta_{\perp}(\alpha)\,r\,.
\end{eqnarray}
Interestingly, beside the non-trivial dependency on the aspect-ratio $\alpha$, the longitudinal componant of the electric field inside the spheroid is a linear function of $x$ only, while the transverse componant is a function of $r$ only. The non-relativistic equations of motion in such an electric field for a particle with charge $Z e$, mass $m$ and initial position $(x_0,r_0)$ simply read:
\begin{eqnarray}
\label{eq_motion_x}\frac{d^2}{dt^2}\tilde{x} = \kappa\,\zeta_{\Vert}(\alpha)\,\tilde{x}\,,\\
\label{eq_motion_r}\frac{d^2}{dt^2}\tilde{r} = \kappa\,\zeta_{\perp}(\alpha)\,\tilde{r}\,,
\end{eqnarray} 
where $\alpha$ and $\kappa = 4\pi k\,\rho_c\,(Z e/m)$ depend only on time, and $\tilde{x}=x/x_0$ and  $\tilde{r}=r/r_0$. Considering initial conditions $\tilde{x}\vert_{t=0} = \tilde{r}\vert_{t=0} = 1$ and $\partial_t\,\tilde{x}\vert_{t=0} = \partial_t\,\tilde{r}\vert_{t=0} = 0$, Eqs.~(\ref{eq_motion_x}) and~(\ref{eq_motion_r}) are found to be independent on the initial coordinates $x_0$ and $r_0$. As a consequence, a particle initially located at a position ($x_0, r_0$) will subsequently be at a position ($x_0\,\tilde{x}, r_0\,\tilde{r}$) where $\tilde{x}$ and $\tilde{r}$ do not dependent on the initial position. Therefore, it is straightforward to obtain that (i) CE of an initially uniformly charged spheroid conserves the spheroidal shape (albeit, as we will see in Sec.~\ref{sec2.2}, with a time-dependent aspect-ratio), and that (ii) the charge distribution inside the spheroid remains uniform (albeit time-dependent).

Although the above calculations have been performed considering Coulomb interaction in a spheroid, we would like to stress that similar conclusions can be drawn for the more general case of an ellipsoid, as well as for any Coulomb-like force, such as, e.g., gravitation~\cite{lin_ApJ_1965}.

The above considerations strongly simplify the modelling of CE of a uniformly charged spheroid. The problem can now be solved by considering equations for the evolution of the longitudinal and transverse radii of the spheroid.

\subsection{Coulomb explosion of a uniformly charged spheroid}\label{sec2.2}

\subsubsection{Governing equations}\label{sec2.2.1}

Let us consider a uniformly charged spheroid with initial radii $w_{\Vert,0}$ and $w_{\perp,0}$, initial ion charge density $n_0$. Obviously, the total charge $Q = (4\pi/3)\,w_{\Vert,0}\,w_{\perp,0}^2\,(Z\,e\,n_0)$ is conserved during explosion. The non-relativistic equations of motion~(\ref{eq_motion_x}) and~(\ref{eq_motion_r}) are also valid for particles initially located on the outer shell of the spheroid at $(x=w_{\Vert,0},r=0)$ and $(x=0,r=w_{\perp,0})$. Then, using Eqs.~(\ref{eq_field_1}) and~(\ref{eq_field_2}), it is straightforward to derive a system of two second order differential equations on the time-dependent longitudinal and transverse radii $w_{\Vert}$ and $w_{\perp}$:
\begin{eqnarray}\label{eq_system1}
\frac{d^2}{dt^2}w_{\Vert} = \omega_{p0}^2\,\frac{w_{\perp,0}^2\,w_{\Vert,0}}{w_{\perp}^2}\,\zeta_{\Vert}\left(\frac{w_{\perp}}{w_{\Vert}}\right)\,,\\\label{eq_system2}
\frac{d^2}{dt^2}w_{\perp} = \omega_{p0}^2\,\frac{w_{\perp,0}^2\,w_{\Vert,0}}{w_{\perp}\,w_{\Vert}}\,\zeta_{\perp}\left(\frac{w_{\perp}}{w_{\Vert}}\right)\,,
\end{eqnarray}    
where we have introduced the plasma frequency:
\begin{eqnarray}\label{eq_wp0}
\omega_{p0}=\sqrt{Z^2 e^2\,n_0/(\epsilon_0\,m)}\,.
\end{eqnarray}
Let us now normalize the time to $\omega_{p0}^{-1}$ ($\tau = \omega_{p0}\,t$), the longitudinal radius to $w_{\Vert,0}$ ($\tilde{w}_{\Vert}=w_{\Vert}/w_{\Vert,0}$) and the transverse one to $w_{\perp,0}$ ($\tilde{w}_{\perp}=w_{\perp}/w_{\perp,0}$). The system of Eqs.~(\ref{eq_system1}) and~(\ref{eq_system2}) then reads:
\begin{eqnarray}\label{eq_systemN1}
\frac{d^2}{d\tau^2}\tilde{w}_{\Vert} = \frac{1}{\tilde{w}_{\perp}^2}\,\zeta_{\Vert}\left(\alpha_0\,\frac{\tilde{w}_{\perp}}{\tilde{w}_{\Vert}}\right)\,,\\\label{eq_systemN2}
\frac{d^2}{d\tau^2}\tilde{w}_{\perp} = \frac{1}{\tilde{w}_{\perp}\,\tilde{w}_{\Vert}}\,\zeta_{\perp}\left(\alpha_0\,\frac{\tilde{w}_{\perp}}{\tilde{w}_{\Vert}}\right)\,,
\end{eqnarray}  
where $\alpha_0 = w_{\perp,0}/w_{\Vert,0}$ is the spheroid initial aspect-ratio. Assuming that all particles in the spheroid have initially no velocity, one has for initial conditions:
\begin{eqnarray}
\label{eq_IC1} \tilde{w}_{\Vert}(\tau=0)=1\,, \tilde{w}_{\perp}(\tau=0)=1\,,\\
\label{eq_IC2} \frac{d}{d\tau}\tilde{w}_{\Vert}(\tau=0)=0\,, \frac{d}{d\tau}\tilde{w}_{\perp}(\tau=0)=0\,.
\end{eqnarray}
Note that Eq.~(\ref{eq_IC2}) also implies that particles have no initial temperature. For bare ion spheroids, this situation arises when electron removal is fast enough for ion heating through electron-ion collisions to be negligible. Under such circonstances, the hypothesis of uniform charge density should also be verified as long as the initial atomic density is uniform. 

Before discussing in more details the solution of this system, we note that, using these normalizations, velocities in the longitudinal and transverse directions are naturally expressed in units of $\omega_{p0}\,w_{\Vert,0}$ and $\omega_{p0}\,w_{\perp,0}$, respectively. Correspondingly, energies in the longitudinal and transverse directions are normalized to $\Epsilon_{\Vert,0} = \Epsilon_0/\alpha_0$ and $\Epsilon_{\perp,0} = \alpha_0\,\Epsilon_0$, respectively, where we have introduced the characteristic energy:
\begin{eqnarray}
\Epsilon_0 = \sqrt{\Epsilon_{\Vert,0}\,\Epsilon_{\perp,0}} = m\,\omega_{p0}^2\,w_{\Vert}\,w_{\perp}\,.
\end{eqnarray}

Due to the complex dependency of the shape-functions on the time-dependent aspect-ratio, no general (for any $\alpha_0$) analytical solution can be obtained and the system of Eqs.~(\ref{eq_systemN1}) and~(\ref{eq_systemN2}) has to be solved numerically. However, analytical solutions of the system can be obtained in the case of the sphere (where the aspect-ratio remains constant in time $\tilde{w}_{\perp}/\tilde{w}_{\Vert} = \alpha_0 = 1$), in the case of an infinitely large disc ($\alpha_0 \rightarrow +\infty$ and considering $\alpha_0\,\tilde{w}_{\perp}/\tilde{w}_{\Vert} \rightarrow +\infty$ for all times), and in the case of an infinitely long cylinder ($\alpha_0 \rightarrow 0$ and considering $\alpha_0\,\tilde{w}_{\perp}/\tilde{w}_{\Vert} \rightarrow 0$ for all times). We present briefly the analytical solutions for these particular cases (Sec.~\ref{sec2.2.2}) before discussing in more details the numerical solutions for arbitrary initial values of $\alpha_0$ (Sec.~\ref{sec2.2.3}).

\subsubsection{Particular cases}\label{sec2.2.2}

\paragraph{Coulomb explosion of a uniformly charged sphere}

Spherical CE ($\alpha_0 = 1$) has been widely studied in the context of many applications, e.g. in cluster physics~\cite{nishihara_NIMA_2001, last_JCP_2004, sakabe_PRA_2004, islam_PRA_2006, kaplan_PRL_2003,kovalev_JETP_2005}. In this case, the system of Eqs.~(\ref{eq_systemN1}) and~(\ref{eq_systemN2}) reduces to a single differential equation on the sphere radius $\tilde{R} = \tilde{w}_{\Vert} = \tilde{w}_{\perp}$:
\begin{eqnarray}\label{eq_sphere1}
\frac{d^2}{d\tau^2}\tilde{R} = \frac{1}{3\,\tilde{R}^2}\,.
\end{eqnarray}
The first integral of Eq.~(\ref{eq_sphere1}) is obtained after multiplying both sides by $\frac{d}{d\tau}\tilde{R}$ and integrating from $0$ to $\tau$. Considering that all particles have initially no velocity, we obtain:
\begin{eqnarray}\label{eq_sphere2}
\frac{1}{2}\,\left(\frac{d}{d\tau}\tilde{R}\right)^2 = \frac{1}{3} \left( 1- \frac{1}{\tilde{R}}\right)\,.
\end{eqnarray}
This equation describes the transformation of potential energy [right-hand-side (rhs) of Eq.~(\ref{eq_sphere2})] to kinetic energy [left-hand-side (lhs) of Eq.~(\ref{eq_sphere2})] for a particle located on the outer shell of the spheroid. In our normalized units, energies $\tilde{\Epsilon}$ are expressed in units of $\Epsilon_{\Vert,0}=\Epsilon_{\perp,0}=\Epsilon_0 = m\,\omega_{p0}^2\,R_0^2$, where $R_0$ is the initial radius of the sphere and $\omega_{p0}$ is given by Eq.~(\ref{eq_wp0}). The final kinetic energy of an ion of the sphere outer shell is therefore:
\begin{eqnarray}\label{eq_E_S}
\Epsilon_S = \Epsilon_0/3\,.
\end{eqnarray}

\begin{figure}
\begin{center}
\includegraphics[width=8cm]{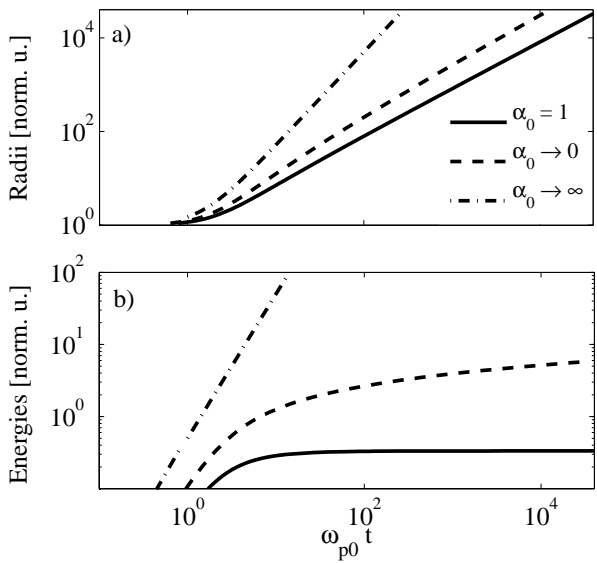}
\caption{Temporal evolution of a)~the normalized spheroid radii ($R/R_0$, $w_{\Vert}/w_{\Vert,0}$ and $w_{\perp}/w_{\perp,0}$), and b)~the maximum energies ($\Epsilon_{max}/\Epsilon_0$, $\Epsilon_{\Vert,max}/\Epsilon_{\Vert,0}$ and $\Epsilon_{\perp,max}/\Epsilon_{\perp,0}$) for spherical explosion (solid curves), cylindrical explosion (dashed curves) and planar explosion (dot-dashed curves), respectively.}
\label{fig_analytics}
\end{center}
\end{figure}

The autonomous differential equation~(\ref{eq_sphere2}) has a formal implicit solution:
\begin{eqnarray}\label{eq_sphere3}
\tau = \sqrt{\frac{3}{2}}\,\int_1^{\tilde{R}} dr \sqrt{\frac{r}{r-1}}\,.
\end{eqnarray}
This solution and the temporal evolution of the outer shell kinetic energy $\Epsilon_{max}$ are shown in Fig.~\ref{fig_analytics}. On long time scales ($\tau=\omega_{p0}\,t > 10$), we find $\tilde{R} \sim \sqrt{2/3}\,\tau$, i.e., most of the potential energy has been transformed into kinetic energy and the sphere expands with the constant velocity $\sqrt{2/3}\,\omega_{p0}\,R_0$. On shorter time scales we actually observe the Coulomb explosion, i.e., $50\,\%$ of the potential energy is transformed in kinetic energy after a time  $\tau \simeq 2.8$, while $80\,\%$ is transformed after $\tau \simeq 7.2$. The characteristic time scale of spherical explosion is therefore the inverse initial plasma frequency $\omega_{p0}^{-1}$.

Following Refs.~\cite{nishihara_NIMA_2001, sakabe_PRA_2004, islam_PRA_2006}, we can derive an analytical expression for the asymptotic ($t \rightarrow \infty$) particle energy distribution. As previously underlined, in our model, the electric field inside the sphere is a linear function of the radius and particles do not overtake each other during expansion. The electric field seen by a particle initially located at $\tilde{r}_0 \le 1$ can thus be easily obtained as a function of the charge $q(\tilde{r}_0)$ inside the sphere with normalized radius $\tilde{r}_0$. The equation of motion for this particle then reads:
\begin{eqnarray}
\frac{d^2}{d\tau^2}\tilde{r} = \frac{\tilde{r}_0^3}{3\,\tilde{r}^2}\,.
\end{eqnarray} 
Then the first integral simply reads (assuming zero initial velocity):
\begin{eqnarray}
\frac{1}{2}\,\left(\frac{d}{d\tau}\tilde{r}\right)^2 = \frac{\tilde{r}_0^3}{3}\,\left(\frac{1}{\tilde{r}_0}-\frac{1}{\tilde{r}} \right).
\end{eqnarray}
This equation once more describes the energy conservation and shows that an ion initially located at a position $\tilde{r}_0$ has obtained, at the end of the acceleration process, a kinetic energy $\tilde{\Epsilon}(\tilde{r}_0)=\tilde{r}_0^2/3$. Now, the normalized radial particle density at initial position $\tilde{r}_0$ is simply:
\begin{eqnarray}
\frac{dN}{d\tilde{r}_0} = 3\,\tilde{r}_0^2\,\theta_H\big(1-\tilde{r}_0\big)\,,
\end{eqnarray}
where $\theta_H$ is the Heaviside function, from which we derive the ion energy distribution at $t \rightarrow \infty$:
\begin{eqnarray}\label{eq_SCE_spectrum}
\frac{dN}{d\tilde{\Epsilon}} = \frac{dN}{d\tilde{r}_0}\,\frac{d\tilde{r_0}}{d\tilde{\Epsilon}} = \frac{9}{2}\,\sqrt{3\,\tilde{\Epsilon}}\,\theta_H\big(1/3-\tilde{\Epsilon}\big)\,.
\end{eqnarray}
Hence, one obtains that the particle energy distribution scales as the square-root of the ion energy up to the maximum energy $\tilde{\Epsilon}^{\infty}=1/3$.

At this point, we want to stress that this asymptotic energy distribution can actually be generalized to all times $t$. It is indeed well-known that CE of a uniformly charged sphere is self-similar and that the velocity distribution inside the sphere increases linearly with the distance to its center. From this we can derive the fraction of particles with an energy below $\Epsilon \le \Epsilon_{max}(t)$: $N(\Epsilon)=[\Epsilon/\Epsilon_{max}(t)]^{3/2}$, where $\Epsilon_{max}(t)$ is the maximum particle energy at time $t$. We finally obtain the time-dependent spectrum:
\begin{eqnarray}\label{eq_SCE_tspectrum}
\frac{dN}{d\Epsilon} = \frac{3}{2}\,\frac{\sqrt{\Epsilon}}{\Epsilon_{max}^{3/2}(t)}\,\theta_H[\Epsilon_{max}(t)-\Epsilon]\,.
\end{eqnarray}

\paragraph{Coulomb explosion of a uniformly charged, infinitely long cylinder}

In the case of a uniformly charged, infinitely long cylinder, the system of Eqs.~(\ref{eq_systemN1}) and~(\ref{eq_systemN2}) reduces to a single differential equation for the cylinder radius $\tilde{w}_{\perp}$:
\begin{eqnarray}\label{eq_cyl1}
\frac{d^2}{d\tau^2}\tilde{w}_{\perp} = \frac{1}{2\,\tilde{w}_{\perp}}\,.
\end{eqnarray}
Once more, the first integral of Eq.~(\ref{eq_cyl1}) describes energy conservation:
\begin{eqnarray}\label{eq_cyl2}
\frac{1}{2}\,\left(\frac{d}{d\tau}\tilde{w}_{\perp}\right)^2 = \frac{1}{2}\,\ln \tilde{w}_{\perp}\,.
\end{eqnarray}
In contrast to the spherical case considered above, the logarithmic potential on the rhs of Eq.~(\ref{eq_cyl2}) goes to infinity for increasing $\tilde{w}_{\perp}$. This unphysical behavior follows from our choice of an infinitely long (thus with infinite total charge) cylinder. As a result, the energy of the outer shell artificially formally diverges. 

Again a formal implicit solution can be obtained for the cylinder radius:
\begin{eqnarray}\label{eq_cyl3}
\tau = \int_1^{\tilde{w}_{\perp}} \frac{dw}{\sqrt{\ln w}}\,.
\end{eqnarray}
The temporal evolution of $\tilde{w}_{\perp}$ and the outer shell energy $\Epsilon_{\perp}$ are presented in Fig.~\ref{fig_analytics}. Expansion occurs with an increasing velocity, and no saturation of the kinetic energy is observed.

\paragraph{Coulomb explosion of a uniformly charged, infinitely large disc}

In the case of a uniformly charged, infinitely large disc, the system of Eqs.~(\ref{eq_systemN1}) and~(\ref{eq_systemN2}) reduces to a single differential equation for the disc thickness $\tilde{w}_{\Vert}$:
\begin{eqnarray}\label{eq_disc1}
\frac{d^2}{d\tau^2}\tilde{w}_{\Vert} = 1\,.
\end{eqnarray}
In this case, the disc thickness increases due to a constant electrostatic field. It simply reads:
\begin{eqnarray}\label{eq_disc2}
\tilde{w}_{\Vert} = \frac{\tau^2}{2}+1\,.
\end{eqnarray}
This result is shown in Fig.~\ref{fig_analytics}. In this case as well, no stationary state is obtained and the kinetic energy increases arbitrarily. 

\subsubsection{Numerical solutions for an arbitrary initial aspect-ratio}\label{sec2.2.3}

As previously discussed, in the general case (for any initial aspect-ratio $\alpha_0$), the system of Eqs.~(\ref{eq_systemN1}) and~(\ref{eq_systemN2}) must be solved numerically. In this paper, it is done using a simple Euler method. Numerical solutions for different values of $\alpha_0$ are now discussed.

\begin{figure}
\begin{center}
\includegraphics[width=8cm]{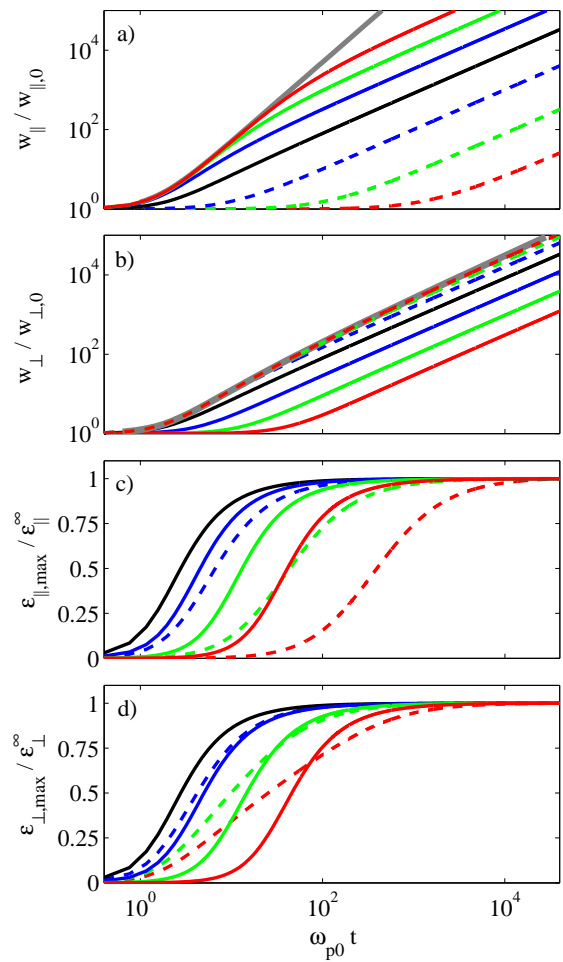}
\caption{Temporal evolution of spheroidal Coulomb explosion: a)~longitudinal radius of the spheroid; b)~transverse radius of the spheroid; c)~energy associated with longitudinal motion; and d)~energy associated with transverse motion. Color codes are as follows: Prolate spheroid (dashed curves, $\alpha_0=10^{-3}$ red, $\alpha_0=10^{-2}$ green and $\alpha_0=10^{-1}$ blue). Sphere (solid black curve, $\alpha_0 = 1$). Oblate spheroid (solid curves, $\alpha_0=10^3$ red, $\alpha_0=10^2$ green and $\alpha_0=10$ blue). The grey solid curves in panels~a) and~b) show analytical predictions for planar ($\alpha_0 \rightarrow \infty$) and cylindrical ($\alpha \rightarrow 0$) explosion, respectively. Note that energies in panels~c) and~d) are normalized to their asymptotic values at $t \rightarrow \infty$.}
\label{fig_we_vs_t}
\end{center}
\end{figure}

\begin{figure}
\begin{center}
\includegraphics[width=8cm]{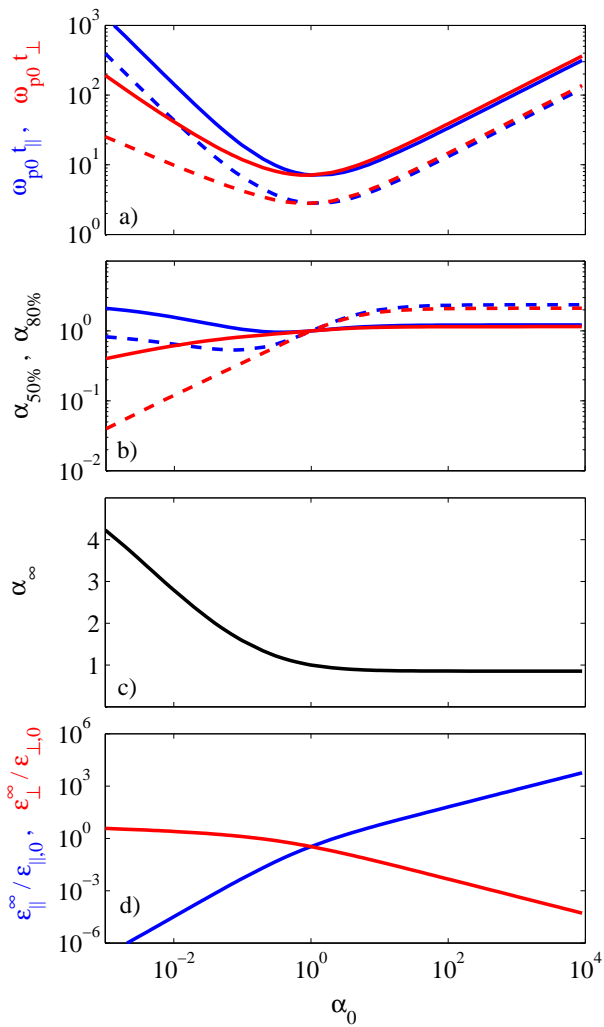}
\caption{Dependence on the initial aspect-ratio $\alpha_0$ of: a)~the time to reach 50\,\% (dashed curves) or 80\,\% (solid curves) of the maximum kinetic energy; b)~the corresponding intermediate aspect-ratio $\alpha_0\,\tilde{w}_{\perp}/\tilde{w}_{\Vert}$; c)~the final aspect-ratio $\alpha_{\infty}$; d)~the final energies $\Epsilon^{\infty}_{\Vert}$ and $\Epsilon^{\infty}_{\perp}$ normalized to $\Epsilon_{\Vert,0}$ and $\Epsilon_{\perp,0}$, respectively. In panels a), b) and d), blue curves account for motion in the longitudinal direction, and red curves for motion in the transverse direction.}
\label{fig_vs_alpha}
\end{center}
\end{figure}

Figure~\ref{fig_we_vs_t} shows the temporal evolution of the longitudinal and transverse radii of the spheroid (Figs.~\ref{fig_we_vs_t}a and~\ref{fig_we_vs_t}b, respectively) and of the longitudinal and transverse kinetic energies (Figs.~\ref{fig_we_vs_t}c and~\ref{fig_we_vs_t}d, respectively). It is complemented by Fig.~\ref{fig_vs_alpha} which shows, as a function of the initial aspect-ratio $\alpha_0$, the times required for the kinetic energy (for purely longitudinal or purely transverse motion) to reach 50~\% or 80~\% of its maximum value (Fig.~\ref{fig_vs_alpha}a), the corresponding aspect-ratio of the spheroid at these times (Fig.~\ref{fig_vs_alpha}b), the final aspect-ratio $\alpha_{\infty}$ (Fig.~\ref{fig_vs_alpha}c) and the final energies normalized to $\Epsilon_{\Vert,0}$ and $\Epsilon_{\perp,0}$ (Fig.~\ref{fig_vs_alpha}d).

Let us first address the case of spherical explosion ($\alpha_0 = 1$). Numerical solutions allow us to recover the analytical findings of Sec.~\ref{sec2.2.2}. The radius and energy evolutions (Figs.~\ref{fig_we_vs_t}) are in perfect agreement with what is presented in Fig.~\ref{fig_analytics}. It is also worth pointing out that for a given initial charge density (i.e., for a fixed value of $\omega_{p0}$), spherical expansion is faster (in terms of energy conversion) than for any other initial aspect-ratio $\alpha_0$. This can be observed in Figs.~\ref{fig_we_vs_t}c and Figs.~\ref{fig_we_vs_t}d, as well as in Fig.~\ref{fig_vs_alpha}a, where the characteristic time to reach a given fraction of the final kinetic energy reaches a minimum for $\alpha_0 = 1$. Finally, we recall that the characteristic time for spherical CE is of the order of $\omega_{p0}^{-1}$ and the final velocities of the outer shell in the longitudinal and transverse directions are $v_{\Vert, max}(t \rightarrow \infty) = v_{\perp,max}(t \rightarrow \infty) = \sqrt{2/3}\,\omega_{p0}\,R_0$. 

In the case of a prolate (cigar-shape, $\alpha_0 \ll 1$) spheroid, expansion occurs mainly in the transverse plane. This intuitive result is illustrated in Figs.~\ref{fig_we_vs_t}a and~\ref{fig_we_vs_t}b where the transverse radius of the spheroid increases much faster than the longitudinal one. Note that, even though the transverse radius of the prolate spheroid evolves initially faster than in the spherical case, spherical expansion remains faster in terms of conversion from potential to kinetic energy (Fig.~\ref{fig_vs_alpha}d). The quasi-stationary state of expansion (i.e., expansion at quasi-constant velocity) is indeed reached later for $\alpha_0 \neq 1$. Note also that saturation in the kinetic energy arises once the spheroid assumes a quasi-spherical shape: when the kinetic energy reaches 80~\% of its maximum value, the spheroid aspect-ratio is indeed quite close to unity (Fig.~\ref{fig_vs_alpha}b). The final aspect-ratio is nevertheless much larger than unity as the final transverse velocity is much larger than the longitudinal one. We also see from Fig.~\ref{fig_vs_alpha}a that spheroidal expansion in the limit $\alpha_0 \ll 1$ occurs on a time scale larger than $(\alpha_0\,\omega_{p0})^{-1}$. Similarly, we see in Fig.~\ref{fig_vs_alpha}d that the asymptotic ($t \rightarrow \infty$) energy for purely longitudinal motion $\Epsilon_{\Vert}^{\infty}$ does not exceed $\alpha_0^2\,\Epsilon_{\Vert, 0} = \alpha_0\,\Epsilon_0$ while the corresponding energy for purely transverse motion $\Epsilon_{\perp}^{\infty}>\Epsilon_0/3$ (for example, for $10^{-5} < \alpha_0 < 10^{-1}$, we find that $\Epsilon_{\perp}^{\infty}$ ranges between $\Epsilon_{\perp,0}=\alpha_0\,\Epsilon_0$ and $7\,\Epsilon_{\perp,0} = 7\,\alpha_0\,\Epsilon_0$). The final transverse energy is therefore significantly larger than the longitudinal one (see also theoretical predictions in Fig.~\ref{fig_comparison}). This is a consequence of the initial geometry, and it is responsible for the final oblate shape of the spheroid observed in Fig.~\ref{fig_vs_alpha}b.

Let us now focus on the case of an oblate (disc-shape, $\alpha_0 \gg 1$) spheroid which is particularly interesting when considering laser-generated ion bunches from a solid target. As expected, expansion initially occurs in the longitudinal direction (Figs.~\ref{fig_we_vs_t}a and~\ref{fig_we_vs_t}b). Transverse expansion eventually occurs later, once the spheroid longitudinal radius becomes comparable to its transverse one. At this time, the spheroid aspect-ratio becomes close to unity and a non-negligible fraction of the potential energy has already been converted into kinetic energy (Fig.~\ref{fig_vs_alpha}b). Finally, one can extract from the numerical results the characteristic expansion time in the limit of large initial aspect-ratio to be $\propto \sqrt{\alpha_0}/\omega_{p0}$. The asymptotic maximum energies can also be easily extracted: $\Epsilon_{\Vert}^{\infty} \sim 0.63\,\alpha_0\,\Epsilon_{\Vert, 0} = 0.63\,\Epsilon_0$ and $\Epsilon_{\perp}^{\infty} \sim 0.46\,\Epsilon_{\perp, 0}/\alpha_0 = 0.46\,\Epsilon_0$. We thus obtain that the final energies in the longitudinal and transverse directions are of the same order. This leads to a final aspect-ratio $\alpha_{\infty} \sim \sqrt{\Epsilon_{\perp}^{\infty}/\Epsilon_{\Vert}^{\infty}} \simeq 0.86$ close to unity, as shown in Fig.~\ref{fig_vs_alpha}c.

\subsection{Energy spectra}\label{2.2.4}
Our model for spheroidal CE allows us to derive the maximum energies for motion along the spheroid principal axes at time $t$. As the density inside the spheroid remains uniform, the velocity distribution along the principal axes has to be a linear function of spatial coordinates:
\begin{eqnarray}
\label{eq_v1} v_{\Vert}(x) &=& (x/w_{\Vert})\,v_{\Vert,max}\,,\\
\label{eq_v2} v_{\perp}(r) &=& (r/w_{\perp})\,v_{\perp,max}\,,
\end{eqnarray}
where $w_{\Vert}$, $w_{\perp}$ are the spheroid radii, and $v_{\Vert,max}$ and $v_{\perp,max}$ the particle maximum velocities at a given time~$t$. Considering an homogeneous charge density in the spheroid, one can easily derive the time-dependent normalized energy distribution for motion along the longitudinal and transverse directions [cf.~Eq.~(\ref{eq_SCE_tspectrum})]:
\begin{eqnarray}
\label{eq_spec1} \frac{dN}{d\Epsilon_{\Vert}}&=&\frac{3}{2}\,\frac{\sqrt{\Epsilon_{\Vert}}}{\Epsilon_{\Vert,max}^{3/2}}\,\theta_H(\Epsilon_{\Vert,max}-\Epsilon_{\Vert})\,,\\
\label{eq_spec2} \frac{dN}{d\Epsilon_{\perp}}&=&\frac{3}{2}\,\frac{\sqrt{\Epsilon_{\perp}}}{\Epsilon_{\perp,max}^{3/2}}\,\theta_H(\Epsilon_{\perp,max}-\Epsilon_{\perp})\,,
\end{eqnarray}
where $\Epsilon_{\Vert,max}=m\,v_{\Vert,max}^2/2$ and $\Epsilon_{\perp,max}=m\,v_{\perp,max}^2/2$ are time-dependent and derived from our model.

\begin{figure}
\begin{center}
\includegraphics[width=8cm]{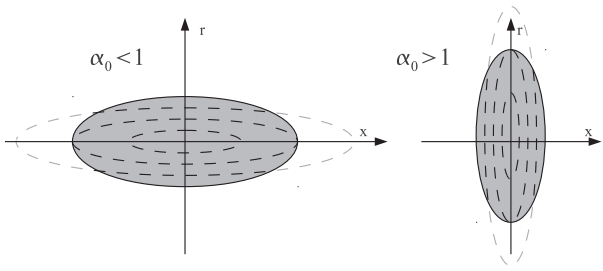}
\caption{Schematic representation of equivelocity surfaces at fixed time $t$ during CE of: (left) a prolate (cigar-shaped, $\alpha_0<1$) spheroid, and (right) an oblate (disk-shaped, $\alpha_0>1$) spheroid. In this two-dimensional representation, equivelocity surfaces correspond to concentric ellipses with the same aspect-ratio.}
\label{fig_isosurf}
\end{center}
\end{figure}

The total energy spectrum can also be derived by considering equivelocity surfaces as concentric homeoids (spheroidal surfaces). We obtain from Eqs.~(\ref{eq_v1}) and~(\ref{eq_v2}) that these homeoids are actually similar, i.e. they have the same aspect-ratio $\alpha_v = (w_{\perp}/w_{\Vert})\,v_{\Vert,max}/v_{\perp,max}$ at fixed time $t$ (see Fig.~\ref{fig_isosurf})~\cite{note1}. This allows us to calculate, for a given energy $\Epsilon$, the fraction $N(\Epsilon)$ of particles in the spheroid with a lower energy, and finally derive the normalized energy distribution.

\begin{widetext}
For a prolate (cigar-shaped, $\alpha_0 < 1$) spheroid, the total energy spectrum reads:
\begin{equation}\label{eq_spec3}
\frac{dN}{d\Epsilon} = \frac{3}{2}\,\left\{\begin{array}{ll}
\frac{\sqrt{\Epsilon/\Epsilon_{\Vert,max}}}{\Epsilon_{\perp,max}} & \,\,{\rm for}\quad \Epsilon<\Epsilon_{\Vert,max}\,,\\
\frac{1}{\Epsilon_{\perp,max}}\,\sqrt{\frac{\Epsilon_{\perp,max}-\Epsilon}{\Epsilon_{\perp,max}-\Epsilon_{\Vert,max}}} & \,\,{\rm for}\quad \Epsilon_{\Vert,max}<\Epsilon<\Epsilon_{\perp,max}\,.
\end{array}\right.
\end{equation}
For an oblate (disk-shaped, $\alpha_0 > 1$) spheroid, the total energy spectrum reads:
\begin{equation}\label{eq_spec4}
\frac{dN}{d\Epsilon} = \frac{3}{2}\,\left\{\begin{array}{ll}
\frac{\sqrt{\Epsilon/\Epsilon_{\Vert,max}}}{\Epsilon_{\perp,max}} & \,\,{\rm for}\quad\Epsilon<\Epsilon_{\perp,max}\,,\\
\frac{\sqrt{\Epsilon/\Epsilon_{\Vert,max}}}{\Epsilon_{\perp,max}}-\frac{1}{\Epsilon_{\perp,max}}\,\sqrt{\frac{\Epsilon-\Epsilon_{\perp,max}}{\Epsilon_{\Vert,max}-\Epsilon_{\perp,max}}} & \,\,{\rm for}\quad \Epsilon_{\perp,max}<\Epsilon<\Epsilon_{\Vert,max}\,.
\end{array}\right.
\end{equation}
In the next Secs.~\ref{sec3} and~\ref{sec4}, we compare these predictions from our model to MD and PIC simulations.
\end{widetext}

\section{Molecular dynamic simulations}\label{sec3}

Molecular dynamics~\cite{frenkel_2002} simulations of CE of an (initially) uniformly charged spheroid are now discussed. To initialize our simulations, $N=5000$ particles [here, hydrogen ions ($Z=1$, $m=1836~m_e$, where $m_e$ is the electron mass)] are randomly placed within a spheroidal volume so that the initial particle density inside this volume is homogeneous. Here we chose an atomic density $n_0 \simeq 9.7 \times 10^{22}~{\rm cm^{-3}}$ (correspondingly, the sphere radius is $R_0 \simeq 2.3~$nm), which is characteristic of hydrogen clusters. To avoid unphysically large contributions to the energy spectrum, we enforce a minimum interparticle distance ($\simeq 75~\%$ of the average interparticle distance $d_{min} \sim n_0^{-1/3}$). Furthermore, all particles were taken initially at rest. Then, knowing the initial state of all particles, we solve Newton's (non-relativistic) equations of motion for each of them using the velocity Verlet scheme~\cite{swope_JPC_1982,allen_1991} and direct calculation of the Coulomb forces between all ions.  

\begin{figure}
\begin{center}
\includegraphics[width=7cm]{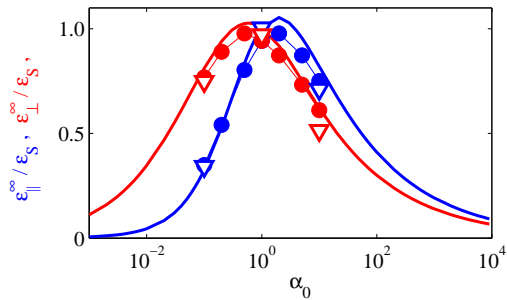}
\caption{Dependence of the maximum energy for the longitudinal and transverse motions (blue and red curves, respectively) on the initial aspect-ratio $\alpha_0$. Solid lines correspond to semi-analytical predictions [Eqs.~(\ref{eq_systemN1}) and~(\ref{eq_systemN2})], circles to MD simulation results, and triangles to PIC simulation results obtained at the end of the simulation. Note that energies are normalized to the asymptotic energy $\Epsilon_S = \Epsilon_0/3$ [Eq.~(\ref{eq_E_S})] obtained for spherical explosion.}
\label{fig_comparison}
\end{center}
\end{figure}

Several simulations were performed only changing the initial aspect-ratio in the range $\alpha_0 = 0.1 - 10$. Figure~\ref{fig_comparison} shows the maximum energy for longitudinal and transverse motion as predicted by our semi-analytical model (Sec.~\ref{sec2}) and as extracted from MD simulations. We stress that, here, energies are normalized to the maximum energy $\Epsilon_S$ [Eq.~(\ref{eq_E_S})] resulting from CE of a sphere with similar density and total charge (in practical units, $\Epsilon_S \simeq 3.1$~keV under current conditions). Figure~\ref{fig_comparison} shows a rather good agreement between our simplified model [solutions of Eqs.~(\ref{eq_systemN1}) and (\ref{eq_systemN2})] and simulations. Also note that MD results confirm the theoretical prediction (clearly shown in Fig.~\ref{fig_comparison}) that, for a given total charge and charge density in the spheroid, the maximum longitudinal (transverse) energy is obtained for a slightly oblate (prolate) spheroid.

\begin{figure}
\begin{center}
\includegraphics[width=8cm]{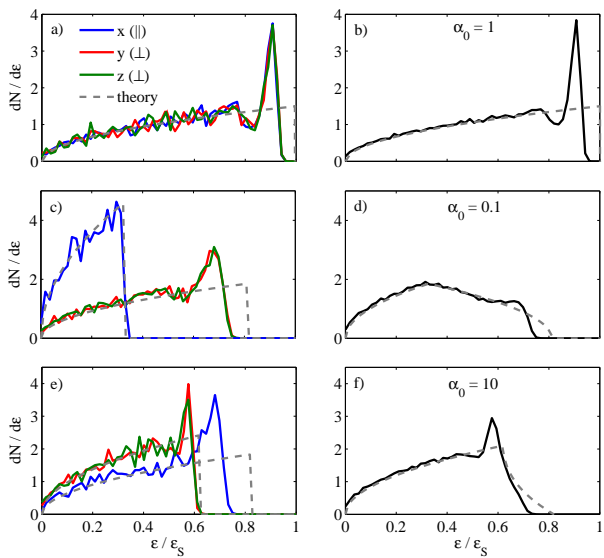}
\caption{Energy spectra obtained from MD simulations at $t \simeq 400~\omega_{p0}^{-1}$: a,b)~for $\alpha_0=1$, c,d)~for $\alpha_0=0.1$, and e,f)~for $\alpha_0=10$. The left panels a,c,e) show the directional spectra: the blue curves correspond to particles emitted within an angle $\pm \pi/20$ of the longitudinal ($x$) direction, the green dashed curves and red curves to particles emitted within an angle $\pm \pi/20$ of the transverse ($y$ and $z$) directions, respectively. The right panels b,d,f) show the total spectra (all particles are accounted for). The gray dashed lines show the theoretical predictions from our model.}
\label{fig_md2}
\end{center}
\end{figure}

To further investigate MD simulation results, we present in Fig.~\ref{fig_md2} the energy spectra obtained at the end of the simulation [at a time $t \simeq 400~\omega_{p0}^{-1}$ (and $\omega_{p0}^{-1} \simeq 2.4$~fs)] and compare them to theoretical predictions from Eqs.~({\ref{eq_SCE_tspectrum}}) and~(\ref{eq_spec1})-(\ref{eq_spec4}). Distributions in longitudinal (transverse) energy are obtained by considering particles emitted within an angle $\theta = \pm \pi/20$ around the longitudinal (transverse) direction. Panels~a) and~b) correspond to spherical CE ($\alpha_0=1$). In this case, the energy distribution is in very good agreement with the theoretical prediction from Eq.~(\ref{eq_SCE_tspectrum}) for energies up to $90~\%$ of the maximum energy. Near the maximum energy, however, a peak appears in the simulation which is not predicted by our model. 

Figures~\ref{fig_md2}c,d and~\ref{fig_md2}e,f show similar results for CE of a prolate (cigar-shape, $\alpha_0=0.1$) spheroid and an oblate (disk-shape, $\alpha_0=10$) spheroid, respectively. Once more, a very good agreement is obtained between theoretical predictions from our model and MD simulations. Only at the maximum energy a peak is observed for any direction of emission in the MD spectra, which is absent in the model spectra. This peak is a consequence of the discrete nature of particles, which leads to a gradual decrease in the particle density at the surface of the spheroid. The thickness of the corresponding surface layer is of the order of the average interparticle separation. The decreasing density results in ion wave breaking (or formation of a shock), whose characteristic signature is a peak in the energy spectrum~\cite{kaplan_PRL_2003,kovalev_JETP_2005,peano_PRL_2005,levy_NJP_2009}. The peak is found to be sensitive to the initial particle distribution and its magnitude decreases with the increase of the total number of particles. This behavior is captured by the MD calculations, which take into account the motion of discrete ions, but is neglected in our simplified model, which considers the evolution of a continuous particle density with a sharp cutoff at the surface.

\section{Particle-in-cell simulations}\label{sec4}

We now present results from simulations of spheroidal CE obtained using the massively parallel 3D PIC code CALDER~\cite{lefebvre_NF_2003}. The PIC simulation technique consists in solving the Maxwell-Vlasov system, and thus offers a mean-field kinetic description for the plasma dynamics~\cite{birdsall_langdon}. In PIC codes, the Vlasov equation is solved by discretizing the particle distribution functions as a sum of so-called macro-particles and by solving, for each of these macro-particles, the corresponding (relativistic) equation of motion in the electromagnetic field. Then, the Maxwell-Amp\`{e}re and Maxwell-Faraday equations are solved on a Yee-mesh using the finite-difference time-domain method~\cite{taflove_2005}. This numerical scheme, coupled to the standard current and charge deposition algorithm in a PIC code, does not automatically satisfy the Poisson equation, which has to be enforced by correcting the electric fields at each time step. In CALDER, this is done by using the usual technique proposed by Boris~\cite{boris_1970}. This study presents the first simulations performed with CALDER in the case of an initially strongly non-neutral plasma for which it is of the utmost importance to accurately correct the electric field at all time-steps. This difficulty can however be alleviated by using charge-conserving algorithms such as the one proposed by Esirkepov~\cite{esirkepov_CPC_2001}. In this case indeed, Poisson equation has to be solved only at the first time-step.

Now, we present simulation results of spheroidal CE for three different values of the initial aspect-ratio $\alpha_0=0.1$, $\alpha_0=1$ and $\alpha_0=10$. All three spheroids consist of fully ionized carbon ions ($Z=6$, $m=12 \times 1836~m_e$) with density $n_0 \simeq 9.2 \times 10^{21}~{\rm cm^{-3}}$ and total charge $Q \simeq 19$~pC. 

\begin{figure}
\begin{center}
\includegraphics[width=8cm]{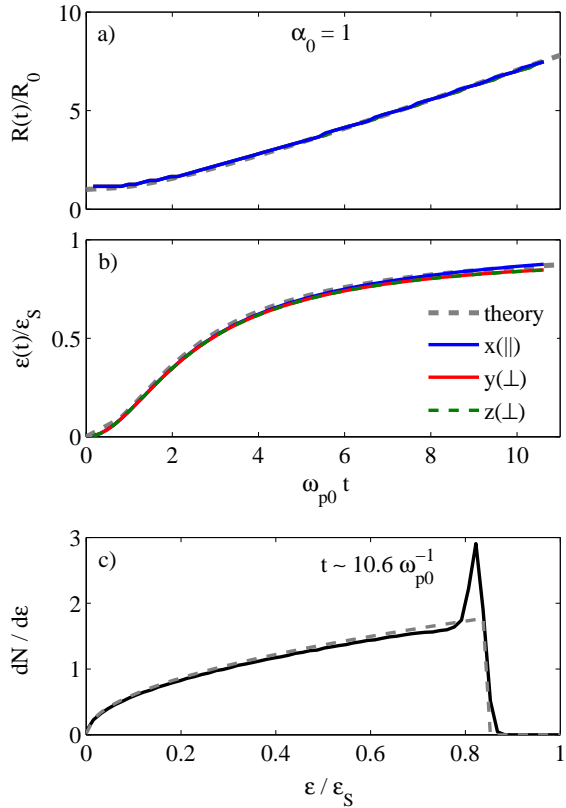}
\caption{Results from PIC simulations for $\alpha_0=1$. a)~Temporal evolution of the sphere radii. b) Temporal evolution of the maximum kinetic energies. c)~Ion energy spectrum at $t \simeq 10.6~\omega_{p0}^{-1}$. Quantities in panels a) and~b) are presented for all three spatial directions: along the $x$-direction (blue), $y$-direction (dashed green) and $z$-direction (red). The gray dashed lines show theoretical predictions from our model.}
\label{fig_pic1}
\end{center}
\end{figure}

\begin{figure}
\begin{center}
\includegraphics[width=8cm]{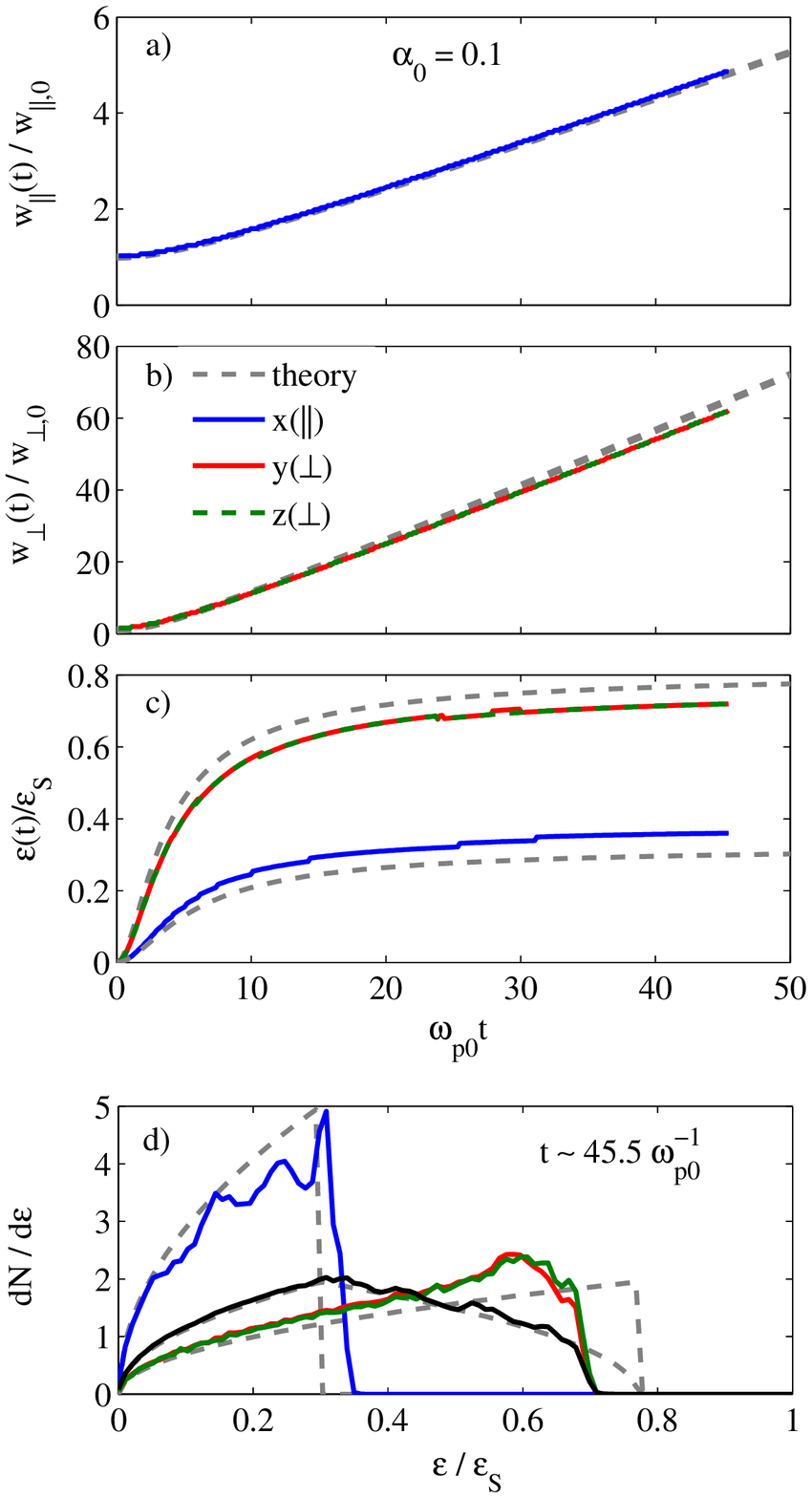}
\caption{Results from PIC simulations for $\alpha_0=0.1$. Temporal evolution of the spheroid radii: a)~in the longitudinal direction, b)~in the transverse direction. c)~Temporal evolution of the maximum energies. d) Corresponding spectra at $t \simeq 45.5~\omega_{p0}^{-1}$, the total spectrum is also shown as a black solid line. Colors codes are chosen as in Fig.~\ref{fig_md2}. The gray dashed lines show theoretical predictions from our model.}
\label{fig_pic2}
\end{center}
\end{figure}

\begin{figure}
\begin{center}
\includegraphics[width=8cm]{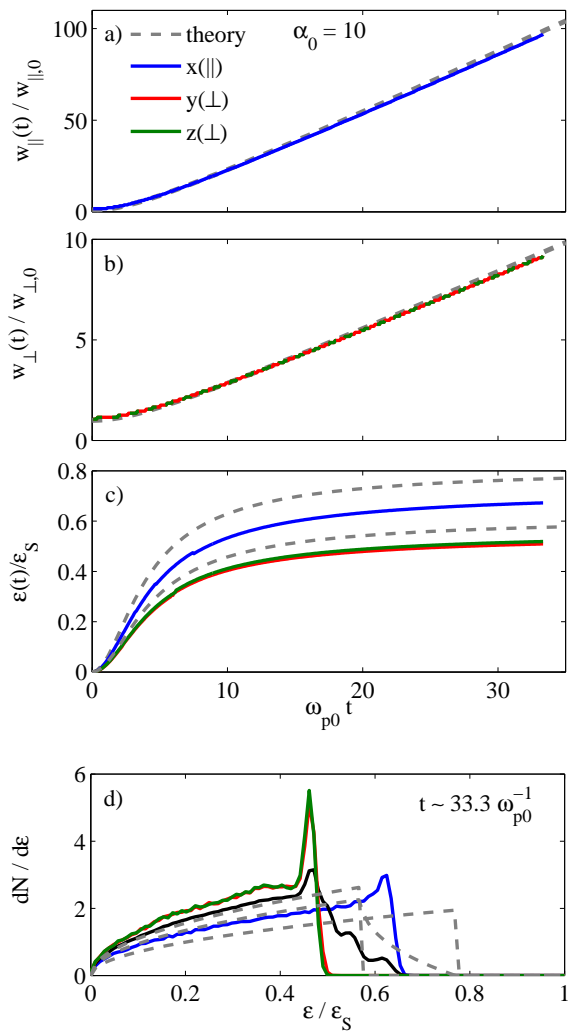}
\caption{Results from PIC simulations for $\alpha_0=10$. Temporal evolution of the spheroid radii: a)~in the longitudinal direction, b)~in the transverse direction. c)~Temporal evolution of the maximum energies. d) Corresponding spectra at $t \simeq 33.3~\omega_{p0}^{-1}$, the total spectrum is also shown as a black solid line. Colors codes are chosen as in Fig.~\ref{fig_md2}. The gray dashed lines show theoretical predictions from our model.}
\label{fig_pic3}
\end{center}
\end{figure}

We first consider CE of a sphere ($\alpha_0 = 1$) with initial radius $R_0 \simeq 80~{\rm nm}$. In this simulation, the mesh size in all three directions is as small as $\Delta x = \Delta y = \Delta z = R_0/20$, and $200$ macro-particles per cell are used. Figures~\ref{fig_pic1}a and~\ref{fig_pic1}b  display the temporal evolution of the spheroid radii (normalized to their initial value $R_0$) and of the maximum kinetic energy [normalized to the final energy predicted from the model $\Epsilon_S$ (in practical units $\Epsilon_S \simeq 12.8$~MeV for present parameters)], respectively. Note that these quantities have been extracted along all three spatial directions ($x$,$y$ and $z$), confirming that the CE dynamics remains spherical within a $\simeq 2~\%$ error. Furthermore, an excellent agreement is found between our theoretical model and simulations concerning the temporal evolution of both the maximum energy and sphere radius. Figure~\ref{fig_pic1}c shows the carbon energy distribution at time $t \simeq 10.6~\omega_{p0}^{-1}$ (for current parameters $\omega_{p0}^{-1} \simeq 4.6~$fs) and their comparison to theoretical prediction from Eq.~(\ref{eq_SCE_tspectrum}). Here again a very good agreement is found between our theoretical model and PIC simulations. It is interesting to see that, in these simulations also, a peak is present at maximum energy. It also originates from a smoothly decreasing ion density at the sphere edge leading to the formation of a shock. In contrast to MD simulations, however, here the decreasing density region is not the result of the discrete particle distribution, but of the projection of the particle density on the mesh. Note also that, at the end of the simulation (at $t \sim 10.6~\omega_{p0}^{-1}$), the maximum ion energy reaches $\simeq 85~\%$ of its theoretical maximum value (for $\omega_{p0}\,t \rightarrow \infty$), which is in good agreement with the dynamics illustrated in Figs.~\ref{fig_we_vs_t}c,d.

Let us now consider the case of a prolate (cigar-shape, $\alpha_0 \ll 1$) spheroid with longitudinal radius $w_{\Vert,0} = 4.6~R_0$ and transverse radii $w_{\perp,0} = 0.46~R_0$ (corresponding to an initial aspect-ratio $\alpha_0 = 0.1$). In this simulation, the mesh size is $\Delta x = \Delta y = \Delta z = R_0/10$ and $500$ particles per cell have been used. Figures~\ref{fig_pic2}a,b,c display the temporal evolution of the normalized spheroid radii and maximum kinetic energies along all three space dimensions, respectively. Figure~\ref{fig_pic2}d shows the energy distributions at the end of the simulation together with their comparison to theoretical predictions from Eqs.~(\ref{eq_spec1}),~(\ref{eq_spec2}) and~(\ref{eq_spec3}). A fair agreement is found between PIC simulations and analytical predictions. 

Finally, the case of an oblate (disk-shape, $\alpha_0 \gg 1$) spheroid with $\alpha_0 = 10$ ($w_{\Vert,0} = R_0/5$ and $w_{\perp,0} = 2\,R_0$) is presented in Fig.~\ref{fig_pic3}. The mesh sizes in this simulation were set to $\Delta x = R_0/20$ and $\Delta y = \Delta z = R_0/10$ and each cell initially contained 300~particles. The temporal evolution of the normalized radii and normalized maximum kinetic energy along the three spatial directions are displayed in Fig.~\ref{fig_pic3}a,b and~\ref{fig_pic3}c, respectively. Figure~\ref{fig_pic3}d shows the energy distributions and their comparison to theoretical predictions from Eqs.~(\ref{eq_spec1}),~(\ref{eq_spec2}) and~(\ref{eq_spec4}). A fair agreement between the PIC simulations and the analytical results is also obtained in this case.

These simulations demonstrate that our simple model correctly describes the CE dynamics of an initially uniformly charged spheroid. We attribute discrepancies between PIC simulation and our model predictions to the limited resolution of the numerical mesh. Due to technical constraints of our computing facilities, we are currently not able to run simulations with a  higher resolution.

\section{Conclusion}\label{sec5}

We have developed a simple, semi-analytical model for CE of a uniformly charged spheroid. In the limit of nonrelativistic particle velocities, this model gives access to the maximum energy a particle can reach at a given time, the time-dependent particle energy distributions, and the characteristic time of CE. All these quantities can be defined as a function of the spheroid aspect-ratio, charge density and total charge. Our theoretical predictions are found to be in remarkably good agreement with particle (both MD and PIC) simulations. 

As 3D kinetic simulations come at a high computational cost, our results are particularly useful when considering acceleration of ions in the pure CE regime, originating from (spherical or non-spherical) clusters, or from thin, solid targets.\\
Indeed, our results should be directly applicable in the so-called CVI regime. This regime where electrons are expelled from the cluster on a time much shorter than the characteristic time of ion motion can be accessed by using either ultra-intense laser or x-ray pulses.\\
Moreover, with the recent progress in nanotechnology, CE of nanostructured targets can be considered. Our results may thus give us simple design guidelines how to optimize target properties, e.g., for inertial fusion applications~\cite{desai_FEC_2000} or to maximize ion collision events for neutron production~\cite{ditmire_NATURE_1999}.\\
Last but not least, our results can also be helpful to model laser-solid target interaction for ion acceleration which is characterized by the emission of short, compact, and highly charged ion bunches. Propagation of these bunches, e.g. through a vacuum, is strongly affected by space charge effects~\cite{grech_LinPA, tikhonchuk_NIMA_2010}. By approximating the accelerated ion bunches as uniformly charged spheroids, the results presented here may allow us to derive the conditions required for limited energy and angular dispersions.

\section*{Acknowledgements}

Numerical simulations were performed using HPC resources at Rechenzentrum Garching and from GENCI at CCRT and CINES (Grant 2010-x2010056304).



\begin{thebibliography}{2}
\bibitem{note1} Note that the aspect-ratio $\alpha_v$ is a function of time.
\end{thebibliography}
\end{document}